\documentclass[letter]{aa}  
\usepackage{graphicx,bm}
\usepackage{txfonts}
\usepackage[colorlinks,citecolor=blue]{hyperref}
\begin{document} 
   \title{Large closed-field corona of WX\,UMa evidenced from radio observations}
   \titlerunning{Large closed-field corona of WX\,UMa}
   \authorrunning{Davis et al.}
   \author{I. Davis \inst{1,2} \and
   H. K. Vedantham \inst{1,3} \and
    J. R. Callingham \inst{1,4} \and
    T. W. Shimwell \inst{1,4} \and 
    A. A. Vidotto \inst{5} \and
    P. Zarka \inst{6} \and
    T. P. Ray \inst{7} \and
    A. Drabent \inst{8}
    }     
    \institute{
ASTRON, Netherlands Institute for Radio Astronomy, Oude Hoogeveensedijk 4, Dwingeloo, 7991 PD, The Netherlands
\and
Department of Physics and Astronomy, University of New Mexico, Albuquerque, USA 
\and
Kapteyn Astronomical Institute, University of Groningen, PO Box 72, 97200 AB, Groningen, The Netherlands
\and
Leiden Observatory, Leiden University, PO Box 9513, 2300 RA, Leiden, The Netherlands
\and
School of Physics, Trinity College Dublin, the University of Dublin, Dublin-2, Ireland
\and
LESIA, CNRS -- Observatoire de Paris, PSL 92190, Meudon, France
\and
Dublin Institute for Advanced Studies, Ireland
\and
Th\"{u}ringer Landessternwarte, Sternwarte 5, 07778 Tautenburg, Germany
}
\date{Received XXX; accepted YYY}

\abstract{
The space-weather conditions that result from  stellar winds significantly impact the habitability of exoplanets.  The conditions can be calculated from first principles if the necessary boundary conditions-- namely on the plasma density in the outer corona and the radial distance at which the plasma forces the closed magnetic field into an open geometry-- are specified. Low frequency radio observations ($\nu \lesssim 200$\,MHz) of plasma and cyclotron emission from stars probe these magneto-ionic conditions. Here we report the detection of low-frequency ($120-167\,{\rm MHz}$) radio emission associated with the dMe6 star WX\,UMa. If the emission originates in WX\,UMa's corona, we show that the closed field regions extends to at least $\approx 10$ stellar radii, that is about a factor of a few larger than the solar value, and possibly to $\gtrsim 20$ stellar radii. Our results suggest that the magnetic-field structure of M dwarfs is in between Sun-like and planet-like configurations, where compact over-dense coronal loops with X-ray emitting plasma co-exist with a large-scale magnetosphere with lower plasma density and closed magnetic geometry.}


   \maketitle
%
\section{Introduction}

M-dwarfs often host multiple terrestrial temperate exoplanets \citep{dressing2015,gillon2017}. Since M-dwarfs display heightened levels of magnetic activity over gigayear timescales, their high energy photon and plasma ejection could adversely affect the habitability of exoplanets \citep{kopparapu2013}.

Magnetohydrodynamical (MHD) simulations of exoplanet space-weather have been used to model the interplanetary plasma density and magnetic field strength around low-mass stars \citep{vidotto2021,vidotto2013,cohen2014,garraffo2017}. These models must necessarily ascribe boundary conditions on the plasma density and velocity, as well as the magnetic field strength and topology on some closed surface. The large-scale magnetic field at the stellar surface can be well constrained by Zeeman Doppler imaging maps \citep{morin2010} but the radial evolution of the field, and the radius at which it transitions from a closed topology to radially distended open topology, cannot be determined without explicit knowledge of the plasma density. The plasma density in these simulations is usually assumed ad-hoc or estimated from the X-ray luminosity of the star. However, the X-ray measurements are likely biased towards over-dense gas that is held confined in short coronal loops (smaller than the stellar radius typically) since X-ray emissivity depends on  the square of the electron density. Instead, it is the lower density plasma that occupies the bulk of the coronal volume that expands as the stellar wind. The density contrast between plasma in short coronal loops and the bulk of the corona could, in principle, be orders of magnitude on M-dwarfs because their intense magnetic fields can support such a pressure difference \citep[see ][ for a theoretical analysis of a generic open-closed field boundary region]{pneuman1968}. As a result, existing MHD simulations have been hampered by a lack of empirically determined boundary conditions for the coronal gas properties.

Polarised lower frequency radio emission, due to the plasma or cyclotron maser mechanism, is expected to originate at much higher coronal layers (many stellar radii as we demonstrate here) instead of from short coronal loops close to the stellar surface where the densest X-ray emitting gas is thought to be confined. Such radio observations can therefore provide a crucial constraint on the density of the outflowing plasma and significantly improve the reliability of exoplanet space-weather simulations.

The vast majority of stellar radio emission has only been studied at frequencies greater than 1\,GHz  \citep{gudel2002} owing mainly to sensitivity  constraints. The advent of sensitive radio telescopes at metre wavelengths has opened up the low-frequency window to stars and brown dwarfs \citep{lynch2017, villadsen2019, vedantham2020,callingham_crdra,elegast}. As part of our ongoing effort to constrain coronal parameters using low-frequency emission, here we focus on WX\,UMa--- a system blindly detected \citep{callingham2020} in the ongoing Low-Frequency Array (LOFAR) Two-metre Sky Survey (LoTSS) \citep{lotss1,lotss2} with the LOFAR telescope \citep{lofar}. We have chosen WX\,UMa from our LoTSS detected stars because its magnetic field strength and topology are well known \citep{morin2010}. Here we use the LOFAR radio data to show that, if the emission originates in the corona of WX\,UMa, then the low frequency of emission, despite the strong magnetic field of WX\,UMa, requires the star to have a closed-field configuration out to a radial distance of $\gtrsim 10$ stellar radii and possibly $\gtrsim 20$ stellar radii. This is substantially larger than radial distances to the open-closed field boundary of $3-5$ stellar radii that has been derived from existing MHD simulations of M-dwarfs stellar winds \citep[][]{vidotto2013,vidotto2014,Kavanagh2021}, but comparable to the size of `slingshot prominences' \citep{cameron1989} proposed to exist on fast-rotating cool stars \citep{jardine2020}. 

\section{Observations \& Results}
\subsection{Low-frequency radio detection}
\begin{figure*}
    \centering
   \includegraphics[width=\linewidth]{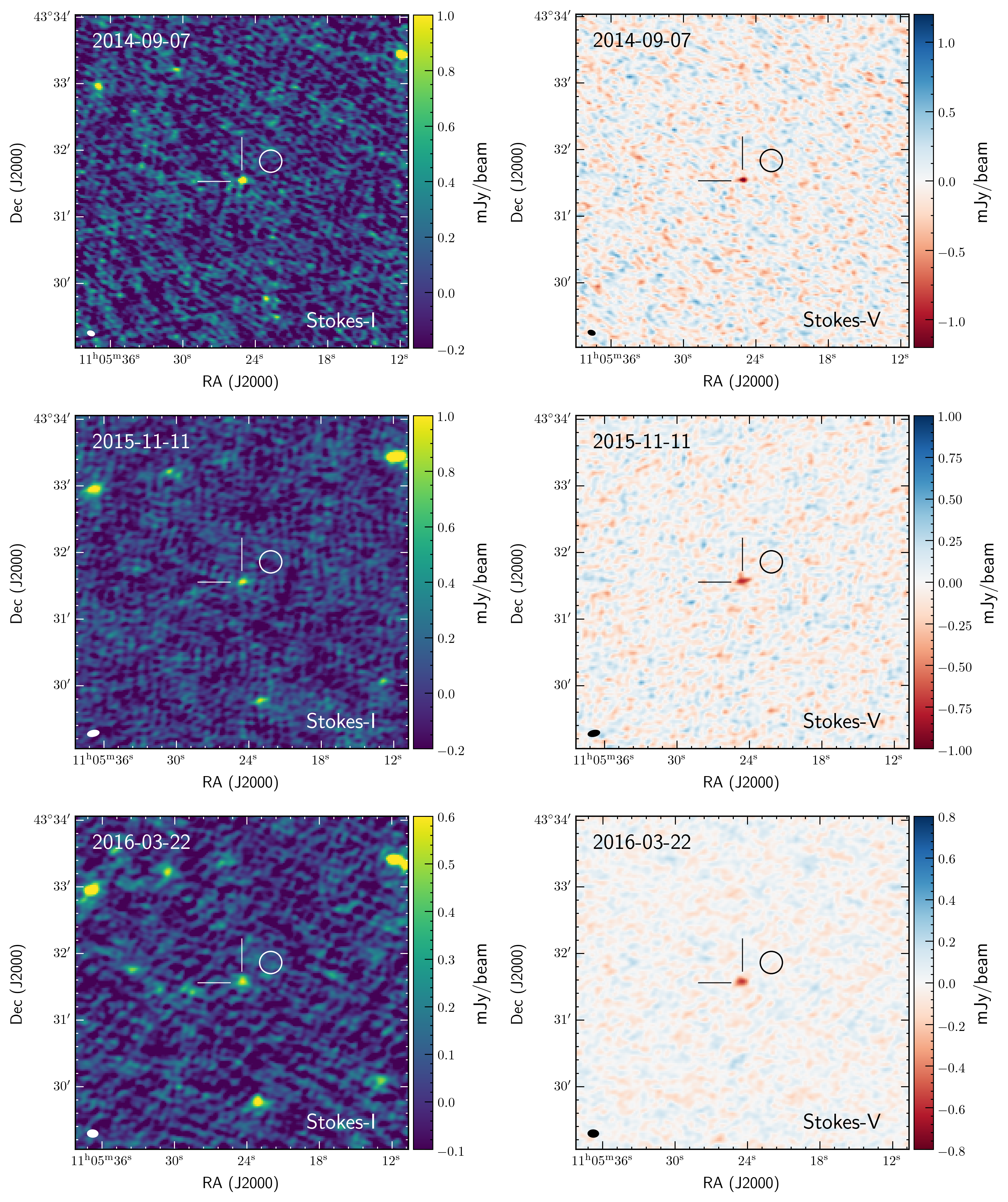}
    \caption{Stokes-I (left) and Stokes-V (right) images of the area surrounding WX\,UMa. The images are formed from the entire 8\,hour observation duration and entire 44\,MHz bandwidth ($120-167\,{\rm MHz}$). The cross-hairs and circles show the proper-motion corrected position of WX\,UMa (GJ412\,B) and GJ412\,A, respectively, using the astrometric data from {\em Gaia} DR2 \citep{gaia}. The flux densities of WX\,UMa and uncertainties are collated in Table \ref{tab:obs}.}
    \label{fig:radio_montage}
\end{figure*}
WX\,UMa was discovered as a radio source in the 120-168\,MHz LoTSS survey \citep{callingham2020} as part of our on-going blind survey for stellar, brown dwarf and planetary radio emission \citep{vedantham2020,vedantham-bdc,callingham2019,callingham_crdra}.
LoTSS surveys the sky with a tessellation of partially overlapping pointings \citep{lotss1,lotss2}. WX\,UMa was observed in four such pointings (see Table \ref{tab:obs} and Fig. \ref{fig:radio_montage}). However, two of the four pointings were observed simultaneously using LOFAR's multi-beaming capability and therefore have redundant information between them. The source is detected in all three independent pointings and has a consistently high ($\gtrsim 70\%$) circular polarisation fraction when averaged over the full $\approx 8$\,h of each observation. Its companion star, GJ\,412\,A that is $\approx 30^{''}$ away, is not detected in any pointing. 
\begin{table*}[]
    \centering
    \begin{tabular}{c c c c c c c} \hline
         {\bf LoTSS field} & {\bf Obs. date} & {\bf Dist.} & {\bf Obs. id} & {\bf Stokes-I (mJy)} & {\bf Stokes-V (mJy)} & {\bf Pol. fraction} \\ \hline
         P5Hetdex & 2014-09-07 & $1^\circ .81$ & 233996 &  $1.8(2)$ & $-1.4(2) $ & $0.8(1) $        \\
         P167+42  & 2016-03-22 & $1^\circ .54$ & 441476 &  $0.6(1)$ & $-0.58(6)$ & $1.0(1)$              \\
         P163+45$^{\ast}$  & 2015-11-11 & $2^\circ . 32$ & 403966 &  $1.1(1)$ & $-0.8(1)$& $0.7(1)$   \\
         P163+42$^{\ast}$  & 2015-11-11 & $2^\circ .31$ & 403966 &  $0.9(1)$ & $-0.6(1)$ &  $0.7(1)$   \\ \hline
    \end{tabular}
    \caption{Details of LOFAR observations of WX\,UMa. Quantities in brackets denote $1\sigma$ root mean square (rms) uncertainty on the least significant digit. WX\,UMa is seen through different gains in the primary beam which leads to differing noise levels. The polarised fraction error is computed by Taylor expanding the ratio to first order. The two exposures with an $^{\ast}$ superscript were taken concurrently using LOFAR's multi-beam capability.}
    \label{tab:obs}
\end{table*}
\subsection{Spectro-temporal properties}
\begin{figure*}
\centering
\includegraphics[width=\linewidth]{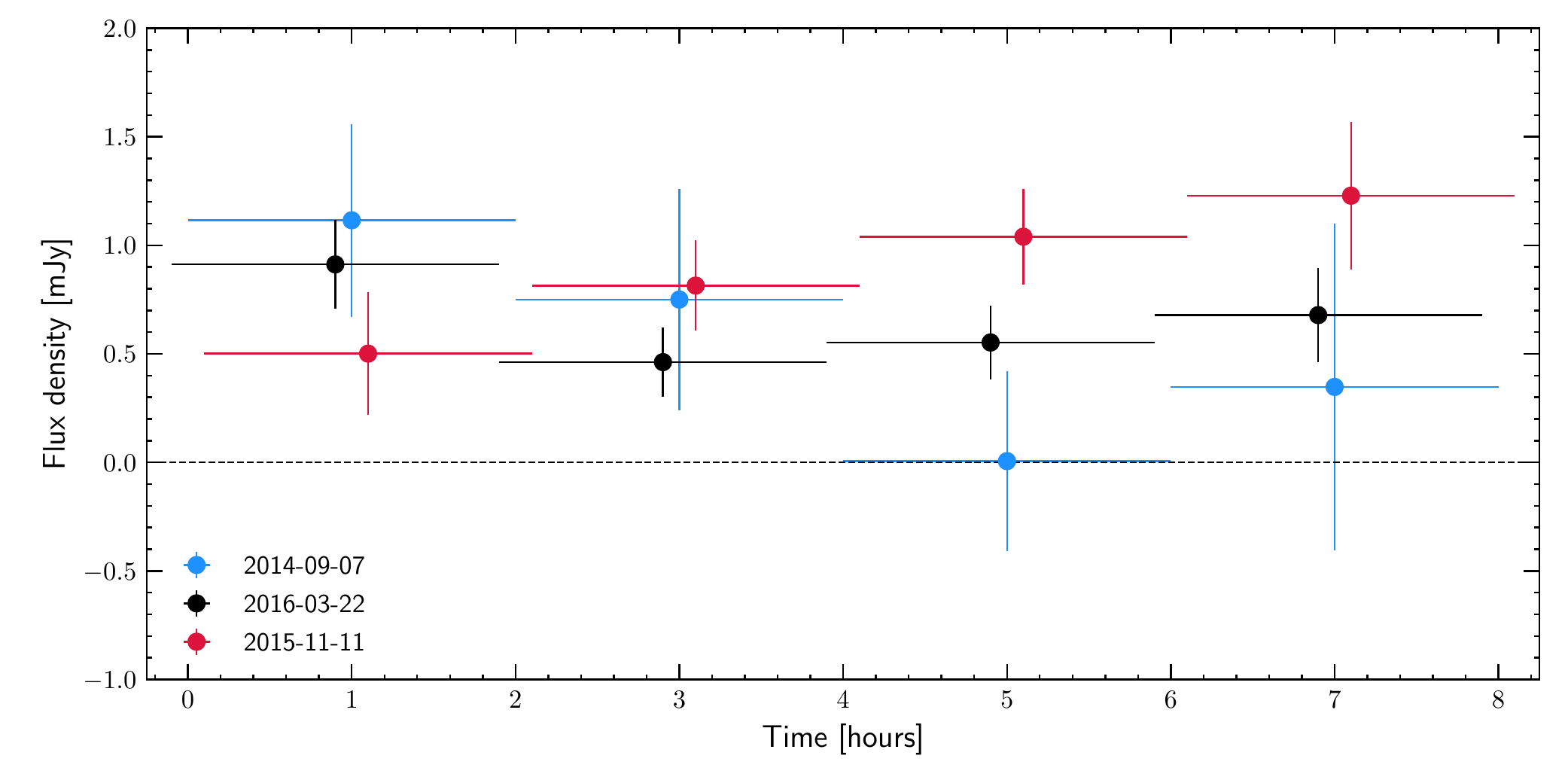}
\caption{Time resolved Stokes-I flux density of WX\,UMa averaged over the full 120-167\,MHz band on the three independent exposures (see Table \ref{tab:obs}). The emission persists of most of the 8\,hr survey pointing duration.\label{fig:lc}}
\end{figure*}

\begin{figure*}
\centering
\includegraphics[width=\linewidth]{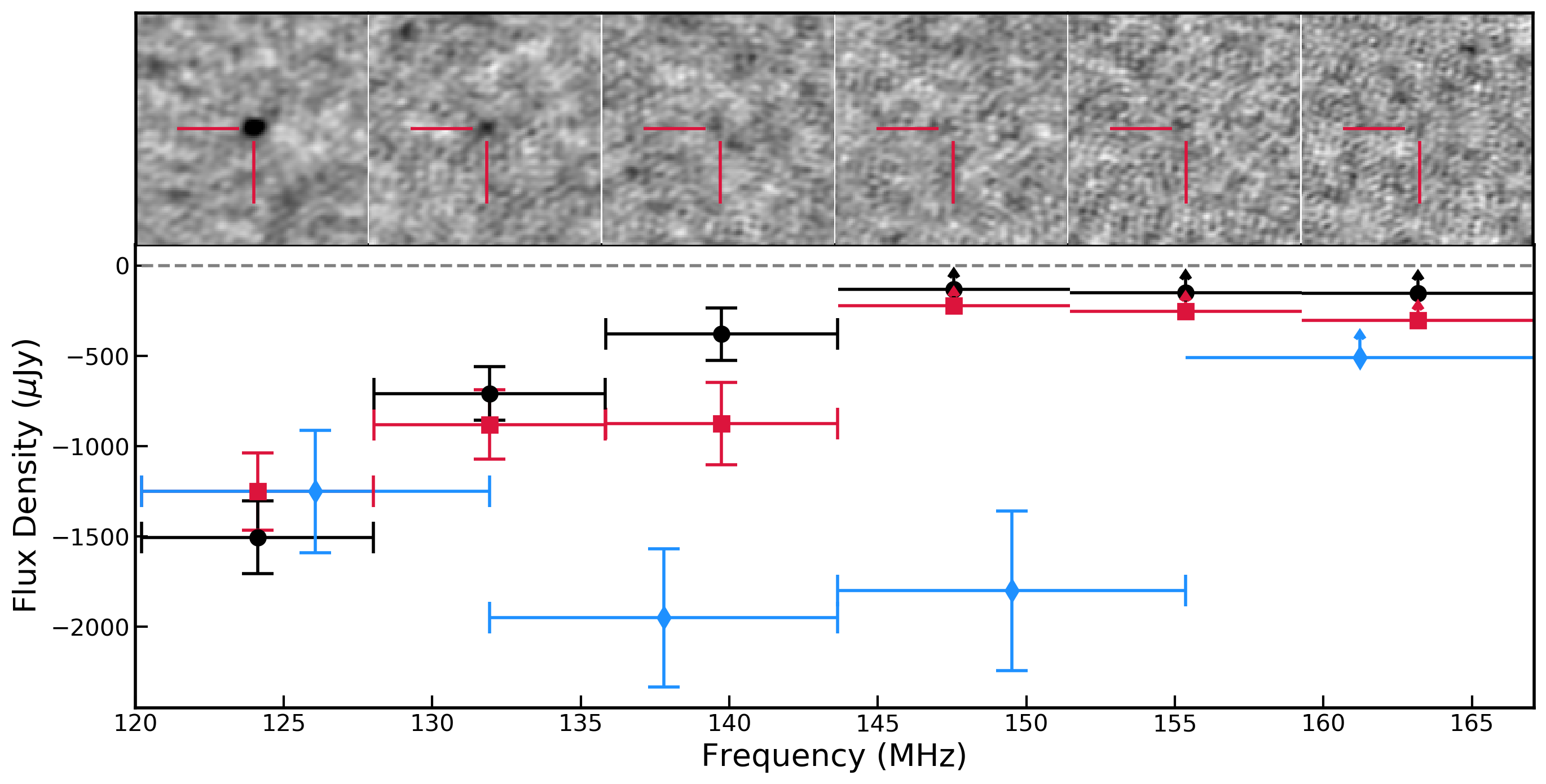}
\caption{Stokes-V spectrum of WX\,UMa in the three independent exposures from the LoTSS survey observations. The three sets of points show the spectrum for the three independent exposures. Upper limits correspond to $1\sigma$. The grayscale images are made from the P167+42 exposure (see Table \ref{tab:obs}) over 8\,hr of synthesis and $\approx 7.8\,{\rm MHz}$ of bandwidth. The red cross-hair mark the position of WX\,UMa. The source consistently shows a drop in flux density magnitude above $\approx 150\,{\rm MHz}$ in all epochs suggesting that this is an enduring property of the emitter. We made a 3.5-$\sigma$ detection of WX\,UMa when averaging over the last three bands in the P167+42 observation, suggesting the spectral evolution at these higher frequencies is a low plateau rather than a cutoff. \label{fig:spec3}}
\end{figure*}

The low frequency radio emission from WX\,UMa is broad-band and persistent. The emission persists for most of the 8\,hr observation block in each exposure, as seen in Fig. \ref{fig:lc}. The spectrum, shown in Fig. \ref{fig:spec3} is particularly interesting, as it has a consistent shape across the observations: it is bright at the low-end of the observational band, and rapidly declines within a few MHz around $\approx 150\,{\rm MHz}$ to a level below our detection threshold. 
For instance, in the 2016-03-22 exposure (P167+42), for which we have the best Stokes-V sensitivity, the emission changes from $\approx-1500$\,$\mu$Jy at 124\,MHz to being undetectable (below 3\,$\sigma$) in the three 7.8-MHz wide channels between $\approx 148$ to $\approx 170\,$MHz, which have an rms noise of $\approx$\,140\,$\mu$Jy. This corresponds to a factor of $\approx 4$ decline over a fractional bandwidth $\Delta\nu / \nu \approx$\,0.15, where $\nu$ is the observing frequency. The decline in other epochs appears to be even more abrupt. However, the available signal-to-noise does not allow us to trace the spectral decline in finer channels. The small number of exposures (three) also do not allow a meaningful search for patterns in the location of the spectral turnover with observation epoch.

We also averaged the images in the three 7.8-MHz channels of the P167+42 that yielded non detections at the $3\,\sigma$ level. The averaged image (not shown here) revealed a faint $3.5\,\sigma$ source at the location of WX\,UMa. This suggests that the emission may settle into a low plateau as opposed to a total cut-off.
Hence, despite the rapid decline, the emission does not appear to `turn-off' entirely at the high frequency end of our band. 

Further details of the radio data processing and extraction of dynamic spectra are given in Appendix \ref{sec:data_proc}.

%
%
%
\section{Discussion}
\begin{table*}
    \centering
    \begin{tabular}{lll}
        \hline
         {\bf Parameter} & {\bf Value} & {\bf Reference/Comments}\\ \hline 
         Spectral class & dM6e & \cite{SpecClass}\\
         Mass ($M_\odot$) & 0.095 & \cite{MassRef} \\
         Radius ($R_\odot$) & 0.12 & \cite{morin2010}\\
         Distance (pc) & 4.9 & \cite{gaia} \\
         ROSAT X-ray luminosity (ergs\,s$^{-1}$) & $3.62 \times 10^{27}$ & \cite{callingham2020}\\
         H$\alpha$ equivalent width ($\textup{\AA}$) & 6.85\,$\pm$\,0.09 & \cite{MassRef}\\
         Surface polar magnetic field$^\ast$ (kG) & 4.53 & \cite{morin2010}\\
         Stellar rotation period (days) & 0.78 & \cite{morin2010} \\
         Average coronal temperature & $5.8\times10^6\,{\rm K}$ & Based on \citet{johnstone2015} \\ Density scale height & $0.63R_\ast$ & Neglecting centrifugal force \\ \hline
    \end{tabular}
    \caption{Known properties of WX\,UMa from literature. $^\ast$\cite{morin2010} also found that the magnetic field is well described by a dipole: ${\bm B} = (B_0/2)R^{-3}(2\cos\theta \bm{ \hat{r}} + \sin\theta \bm{ \hat{\theta}})$, where $B_0$ is the polar surface field strength, $R$ is the radial distance in units of the stellar radius, $\theta$ is the polar angle, and $\bm{\hat{r}}$ and $\bm{\hat{\theta}}$ are the unit vectors along the radial and polar-angular directions.}
    \label{tab:lit_values}
\end{table*}
The known properties of WX\,UMa which will aid in our interpretation of the radio emission are summarised in Table \ref{tab:lit_values}. Because the emission is persistent, we only consider cases where the emitting plasma is trapped within a closed magnetic geometry. This is because plasma that unbinds from the star is expected to expand on a timescale that is much shorter than the duration of our observations. For example, solar radio emission from plasma flowing along open field lines typically lasts seconds to minutes as seen in solar Type-II and Type-III bursts \citep{wild1972}. 

Magnetically trapped plasma in stellar coronae can emit via different mechanisms depending on the energy of the emitting electrons and the ambient magneto-ionic parameters \citep{dulk1985}. The high circular polarisation fraction rules out all known incoherent emission mechanisms. Coherent emission mechanisms that can generate up to $100\%$ circularly polarised emission are fundamental plasma emission and fundamental and second harmonic cyclotron emission. While cyclotron emission occurs at the harmonics of the electron cyclotron frequency \citep{treumann,melrose1982}, plasma emission in a magnetic trap is thought to occur at the harmonics of the upper hybrid frequency \citep{zaitsev1983,stepanov2001} which is the quadrature sum of the plasma and electron cyclotron frequencies.  Both of these mechanisms can account for the long duration of the observed emission. Plasma emission will persist for as long as the turbulent Langmuir wave spectrum that powers the emission is maintained. CMI from ultracool dwarfs is commonly seen to be pulsed at the rotation period due to beaming, but for conducive geometries, it can be visible at all rotation phases. CMI's beaming geometry is discussed in more detail in \S\ref{subsec:SpectralEvolution} and Appendix \ref{app:SpectralEvolution}.

\subsection{Emission site}
\label{sec:emission-site}
The magnetic field of WX\,UMa is largely dipolar with a polar surface magnetic field strength of $\approx 4.5\,{\rm kG}$ \citep{morin2010}. This yields a maximum cyclotron frequency at the surface of $\approx 12.6\,{\rm GHz}$. Because a dipolar field evolves with radial distance $R$ as $R^{-3}$, fundamental cyclotron emission in the observed frequency band can only originate at a radial distance of $\approx 4.7R_\ast$, where $R_\ast$ is the stellar radius.

If the emission is due to the plasma mechanism, the emission site cannot be accurately pinpointed because the plasma density profile in WX\,UMa is not known apriori. However, a lower bound on the radial distance of the emission site can be placed based on the requirement for radiation escape. Escaping radiation in the presence of cyclotron damping is obtained when the ambient cyclotron frequency is sufficiently low: $\nu_c<0.26\nu_p$ (see Appendix \ref{app:emission_mechanism}). Observable plasma emission at $120\,{\rm MHz}$ therefore requires the ambient magnetic field to be below $\approx 11\,$G. For a dipolar field that evolves with radial distance as $R^{-3}$, this value can only be reached on WX\,UMa for $R>7.4\,R_\ast$.
 We note that although there is a narrow window of low cyclotron opacity at small angles to the magnetic field  for $\nu_c>0.26\nu_p$ \citep[see Fig. A1 of ][]{vedantham2020b}, the bulk of the radiation will still be absorbed since spontaneous plasma emission is quasi-isotropic.

\subsection{Polarisation parity}
\label{sec:pol-parity}
The Zeeman doppler imaging (ZDI) maps of WX\,UMa show the star to have an intermediate inclination with a small magnetic obliquity and the south magnetic pole pointed towards us \citep[i.e. field lines pointed towards the star; ][]{morin2010}. 
The observed radio emission is right-hand circularly polarised (see Appendix \ref{app:sign_convention} for details of sign convention). It is therefore in the $o$-mode if it originates in the polar regions (where the field points away from us) and $x$ mode if it originates in the equatorial regions (where the field points towards us). Fundamental plasma emission is polarised in the $o$-mode, whereas cyclotron maser emission can be polarised in the $x$-mode or $o$-mode depending on the ambient plasma density \citep{dulk1985,vedantham2020b}. Because WX\,UMa has a mostly axisymmetric dipolar field \citep{morin2010}, we use cyclotron maser emission from Jupiter as a benchmark for our discussion for the possibility of cyclotron maser emission from WX\,UMa \citep{zarka1998}.  Following this, the unstable electron distribution in a closed field region necessary to drive the cyclotron maser is expected to be situated close to the magnetic poles, and therefore, we only consider $o$-mode cyclotron emission. The $o$-mode cyclotron maser for a loss-cone distribution is expected to dominate for $0.1\lesssim \nu_c/\nu_p \ll 1$, with the $x$ mode dominating at lower plasma densities ($\nu_c/\nu_p\lesssim 0.1$) \citep{hewitt1982, melrose1982}.
In summary, the observed polarity requires the emission to originate in the polar regions and be driven by either fundamental plasma emission, or $o$-mode cyclotron maser emission in which case $\nu_c/\nu_p\gtrsim 0.1$.
\subsection{Extent of the closed field region}
To recap, the two possibilities for the emission mechanism are: (a) plasma emission at a radial distance of $>7.4R_\ast$ with ambient density of  $\approx 2\times 10^8\,{\rm cm}^{-3}$, and (b) cyclotron maser at a radial distance of $4.7R_\ast$ with ambient density of $\gtrsim 2\times 10^6{\rm cm}^{-3}$. Detailed MHD simulations are necessary to use these constraints to determine the size of the closed field region accurately, but we can place preliminary constraints with simplifying assumptions. Let us assume that the density has a hydrostatic profile with the inclusion of the centrifugal force \citep[see eqn. 3 of ][]{havnes1984}. We then use the dipolar structure of the magnetic field and compute the equatorial distance at which the gas pressure ($nk_BT$ where $T$ is the temperature, $n$ is the number density and $k_B$ is Boltzmann's constant), is equal to the magnetic pressure. Beyond this radial distance, we assume that the gas forces open the field lines. So we identify this radial distance as the radius of the closed field region. Using this procedure and values relevant for WX UMa from table \ref{tab:lit_values} we find that the closed-open boundary occurs at radii of $\approx 10R_\ast$ and $22R_\ast$ for the case of fundamental plasma emission and cyclotron emission respectively. Finally, the co-rotation radius, $R_{\rm C} = (GM/\Omega^2)^{1/3}$ ($G$ is the gravitational constant, $\Omega$ is the angular speed, $M$ is the stellar mass) is about $13.6 R_\ast$ for WX UMa. It is noteworthy that the corotation radius is comparable to the approximate open-closed boundary for the case of plasma emission. 

\subsection{Source of emitting electrons}
With these considerations, it is still presently unclear what the source of the emitting electrons is in WX\,UMa, but we can draw analogies with two known systems. We first consider the solar paradigm for the case of plasma emission. In this case, the ultimate energy source is thought to be magnetic reconnection in coronal loops close to the surface \citep{priest2002}. However, the brightness temperature arguments in Sec. \ref{sec:bright-temp} show that the solar paradigm can only apply to WX\,UMa if its entire coronal volume out to a radial distance of $\gtrsim 7.4R_\ast$ is filled with flaring loops which is unprecedented. We may alternatively adopt the Jovian paradigm, where cyclotron maser emission is emitted by suprathermal electrons \citep{dulk1985} that are accelerated by breakdown of co-rotation between the magnetic field and the magnetospheric plasma far from the planet's surface, or by unipolar induction by its moons \citep{goldreich1969}. A similar acceleration mechanism sited in the equatorial plasma sheet at few tens of stellar radii may be responsible for WX\,UMa's radio emission. Such an acceleration mechanism has also been suggested for magnetic chemically peculiar stars \citep{linsky1992}. It it also noteworthy that both WX\,UMa and magnetic chemically peculiar stars have strong large scale magnetic fields, hot X-ray coronae and emit coherent circularly polarised emission \citep[e.g. CU Vir, ][]{triglio2000}.

\subsection{Brightness temperature}
\label{sec:bright-temp}
The brightness temperature of the radio emitter assuming isotropic emission is  $T_b\approx 10^{12.5}x^{-2}\,{\rm K}$ where $x$ is the radius of the emitter in units of the stellar radius. 
Theory allows the brightness temperature of cyclotron maser emission to be many orders of magnitude higher than this value \citep{treumann,melrose1982}, and hence, the observed value can be readily accommodated.
Fundamental plasma emission on the other hand cannot attain brightness temperatures $\gg 10^{12}\,{\rm K}$, especially at such low frequencies \citep{vedantham2020b}. As shown in \S\ref{sec:emission-site}, the projected source radius for plasma emission is $x\approx 7.4$, giving an observed brightness temperature of $\sim 10^{11}\,{\rm K}$.  Although we note there is no precedence for plasma emission to be simultaneously occurring at this brightness throughout the corona (at $x = 7.4$ in this case), such a brightness temperature can be reached by plasma emission based on canonical theory if the plasma is highly turbulent (see Appendix \ref{app:sign_convention}). Therefore, brightness temperature arguments do not provide a clear discriminant between the two plausible emission mechanisms.

\subsection{Origin of the spectral evolution}\label{subsec:SpectralEvolution}
Coherent stellar and planetary radio emission are known to display narrowband spectral features but their frequency centroid generally varies stochastically with time \citep[see ][ for examples]{litvinenko2009,villadsen2019}.\footnote{A possible exception is AD\,Leo whose bursts observed by \citet{villadsen2019} show an apparent cut-off near-consistently around 1.5\,GHz.} As such, the rapid spectral evolution from WX\,UMa is not unusual, but the persistent location of the spectral turn-over is noteworthy. The persistence suggests that it encodes meaningful information on the nature and geometry of emission. 

The expressions for the emissivity and absorption coefficients of various emission processes themselves vary smoothly, to polynomial order, with frequency \citep{1996ASSL..204.....Z}. We must therefore seek an explanation for the rapid spectral evolution in terms of beaming, resonance conditions or abrupt boundaries over which the plasma properties can change substantially. With this in mind, we consider three plausible explanations.
\begin{itemize}
    \item {\em Beaming:} If the emission is driven by the cyclotron maser mechanism, then a natural explanation for the cut-off would be due to beaming of the emission. Cyclotron maser emission is beamed into a hollow cone with a large opening angle, $\theta\lesssim \pi/2$, small thickness, $\Delta\theta\ll\theta$ and whose axis is parallel to the ambient magnetic field. Radiation cones axes on a given field line make smaller angles with the magnetic axis at larger frequencies as illustrated in Fig. \ref{fig:ecmi_geometry}. Hence it is possible to construct geometries where the emission goes out of sight at higher frequencies. For example, consider emission originating on field lines that intersect the equatorial plane at some radius $L$. If the magnetic field line makes a polar angle $\alpha(\nu)$ at the point where emission at frequency $\nu$ originates, then the emission will be visible to all observers within a magnetic co-latitude `visibility band' that lies between $\theta-\alpha(\nu)-\Delta\theta$ and $\theta+\alpha(\nu)+\Delta\theta$. If $\alpha(\nu)$ evolves sufficiently rapidly with $\nu$, then a spectral cut off can be explained by beaming. Could a spectral evolution be observed due to the visibility-band edge moving over the line of sight to the Earth within the observed frequency range? The function $\alpha(\nu)$ can be computed for a given $L$, surface magnetic field strength and dipolar geometry. For the case of LOFAR-band emission from WX\,UMa, the slope $\partial\alpha(\nu)/\partial \nu$ varies between $0^\circ.05\,{\rm MHz}^{-1}$ and $0^\circ.12\,{\rm MHz}^{-1}$ as $L$ is varied between $20R_\ast$ and $6R_\ast$ (further details in Appendix).  This corresponds to a shift in the visibility band by $1^\circ$ and $2^\circ.4$ within a 20\,MHz band within the observed frequency range. The angular shift is significantly smaller than the characteristic width of the emission cone's surface, $\Delta\theta$ \citep{zarka1998,melrose1982} as well as other geometric parameters such as magnetic obliquity. Hence, an apparent cutoff due to beaming is possible but with a somewhat contrived geometric setup. Nevertheless, the beaming hypothesis can be tested by a search for a characteristic rotation and spectral modulation of the radio light curves at the rotation rate of WX\,UMa, which is an expected result of any magnetic obliquity in the system. 
    \begin{figure}
        \centering
        \includegraphics[width=\linewidth]{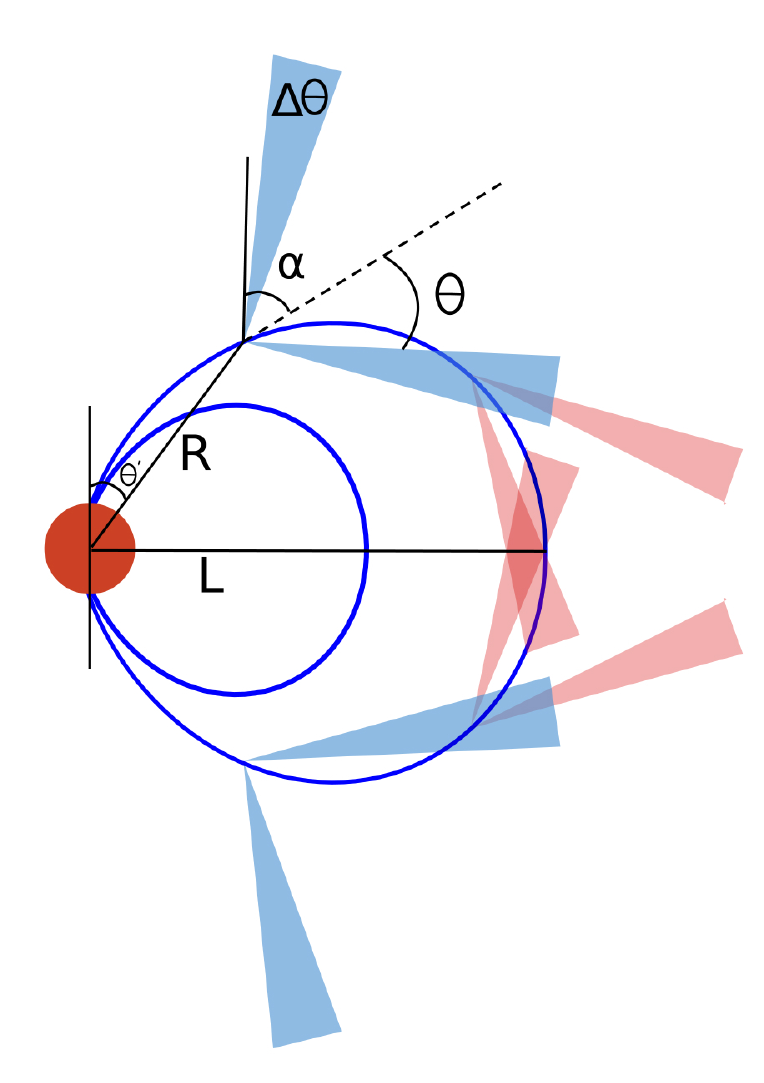}
        \caption{A not-to-scale sketch showing the geometry of cyclotron maser emission on a single dipolar field line that lies in the plane of the page. The higher frequency emission cone (blue) is oriented at a smaller polar angle, $\alpha$, when compared to the emission cone at a lower frequency (red). Here, $L$ is the distance to the field line for a polar angle of $0^{\circ}$, $R$ is the distance from the star's centre to the emission cone along the field line,  $\theta$ is the angle between the direction of emission and the interior wall of the cone, and $\Delta\theta$ is the angular width of the emission cone's wall. The effects of $\alpha$ and $\theta$ on emission frequency has been exaggerated here and is not to scale for the LOFAR frequency range. \label{fig:ecmi_geometry}}
    \end{figure}
    \item {\em Resonant cyclotron absorption:} If the emission is due to fundamental plasma emission, then the cut-off could be due to resonant cyclotron absorption of higher frequency emission that originates deeper in the corona. This is possible because the cyclotron frequency evolves with radial distance as $R^{-3}$ whereas the plasma frequency can evolve much more gradually with radial distance, particularly if the density scale height is large. In such a situation, it is possible for the radiation escape condition to be satisfied at larger radii where lower frequency emission originates, but not at lower radii, where higher frequency emission originates, leading to a cut-off. Fig. \ref{fig:cycl-abs} shows plausible hydrostatic density profiles along different field lines, each with a different base density that leads to catastrophic cyclotron absorption within the LOFAR band. The density profiles all require that the base coronal density is restricted to a narrow range of values (about a 10\% range) around $10^{9.1}\,{\rm cm}^{-3}$ which is comparable to the solar value. If this is the correct mechanism, then similar emission (long duration, highly polarised) must not be visible from WX\,UMa at frequencies above the LOFAR band and long term monitoring should not show clear signs of rotation modulation in the LOFAR-band intensity. We note that WX UMa is detected between 2 and 4 GHz in the ongoing VLASS survey \citep{vlass} at a flux density of $1.3(1)$\,mJy.\footnote{Based on the VLASS `quick-look' images. See https://science.nrao.edu/science/surveys/vlass} However, the VLASS survey observes with 3\,minute exposures which is not long enough to associate the VLASS-detected emission with the long duration LOFAR-detected emission as opposed to a shorter duration burst.
    \begin{figure}
        \centering
        \includegraphics[width=\linewidth]{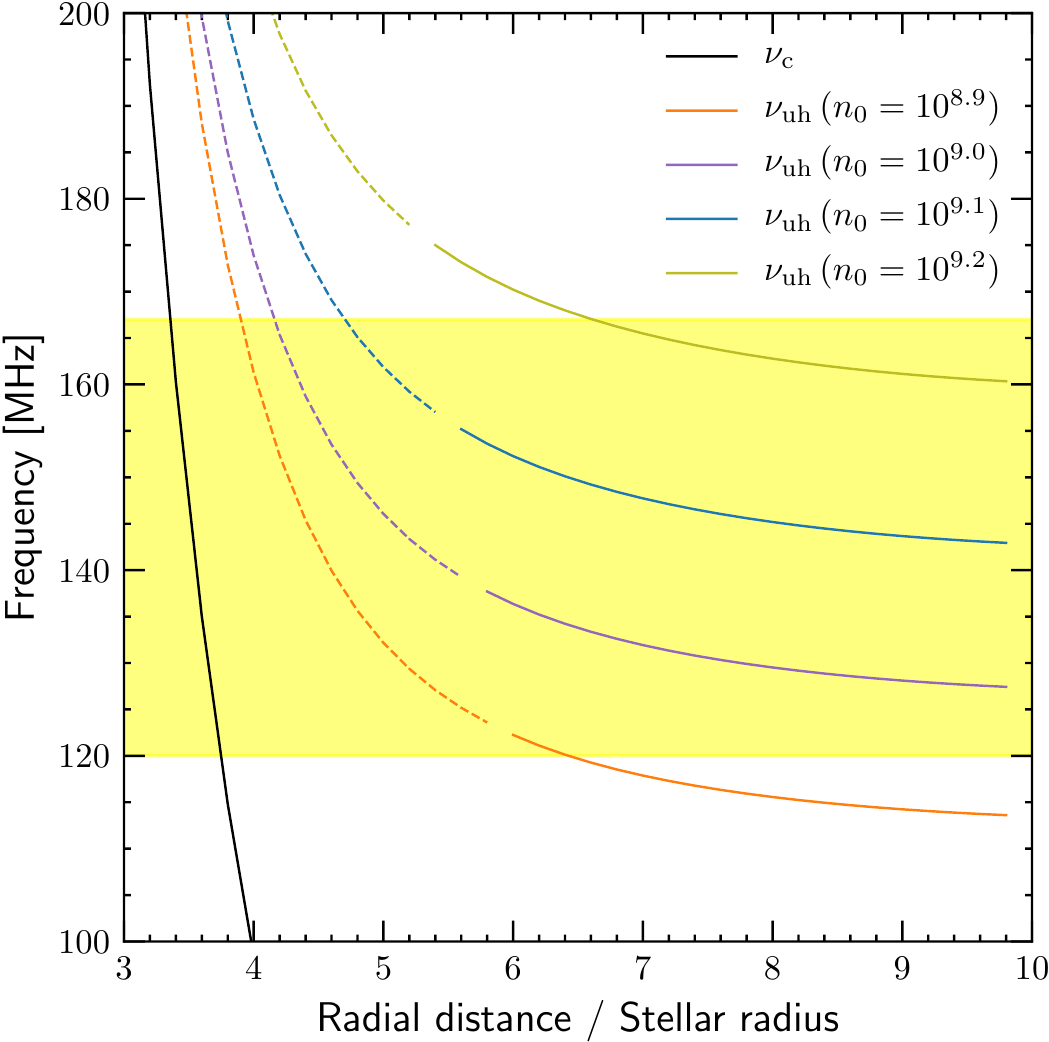}
        \caption{Plot of the radial profile of the cyclotron frequency (solid black line) and the upper hybrid plasma frequency (coloured lines) for different base plasma density values given in units of $cm^{-3}$. The plasma has been assumed to follow a hydrostatic density structure along any given field-line while taking into account centrifugal force \cite[their equation 3]{havnes1984}. The broken lines denote regions where cyclotron absorption will preclude radiation escape. The yellow-band shows the observed frequency range.}
        \label{fig:cycl-abs}
    \end{figure}
    \item {\em Surface cut-off}: The rapid spectral evolution can be explained if the bulk of the emission does not originate in WX\,UMa's corona but instead in the magnetosphere of a closeby (unresolved) sub-stellar companion and the cut-off frequency corresponds to a cyclotron frequency consistent with a surface magnetic field strength of $\sim40$\,G, much in the same way Jovian cyclotron emission cuts-off at around 40 MHz. This would be an unprecedented conclusion if true, as it would amount to a radio discovery of a sub-stellar companion (exoplanet or brown dwarf) around WX\,UMa. A prediction of this interpretation is periodic rotational modulation of the radio emission at a rate that does not match the known rotation rate of WX\,UMa. However, this property cannot be inferred from the current data since the emission has been detected in all three independent pointings.

\end{itemize}
In summary, we have identified three possible mechanisms to explain the rapid spectral evolution all of which can be tested with monitoring the LOFAR-band emission over several rotation cycles and with gigahertz frequency observations.

\section{Summary and outlook}
We have detected persistent highly circularly polarised ($\gtrsim 70\%$) radio emission around 144\,MHz from WX\,UMa. If the emission originates from WX\,UMa's corona, there are two plausible emission mechanisms that can account for the radio emission: (a) plasma emission at the fundamental at a radial distance of $\gtrsim 7.4R_\ast$ where the ambient plasma density is $\approx 10^{8.5}\,{\rm cm}^{-3}$, and (b) $o$-mode cyclotron maser emission in which case the emission originates at a radial distance of $\approx 4.7R_\ast$ where the ambient coronal density is $10^{6.5}\,{\rm cm}^{-3}$.

We expect our plasma density constraints to provide guidance to MHD simulations of the stellar wind of WX\,UMa. A simple balance between gas and magnetic pressure yield the radial extant of the closed field regions to be  $\approx 10R_\ast$ (plasma emission) and $\approx 22R_\ast$ (cyclotron emission) respectively. Such radii are substantially larger than those that have been inferred from several MHD wind simulations of M-dwarfs \citep[see for e.g. ][]{vidotto2014}, but are comparable to recent simulation of M-dwarf winds with low mass-loss rates \citep{Kavanagh2021}. Interestingly, the closed$-$open field boundary in the equatorial plane for the plasma emission model is close to the co-rotation radius for WX\,UMa which is the anticipated extent of slingshot prominences \citep{jardine2020}.

WX\,UMa's radio spectrum shows a rapid decline within the observing band which could be a result of beaming of cyclotron emission or due to cyclotron absorption of plasma emission. Alternatively, the radio emission we observe might have originated not on WX\,UMa but on a sub-stellar companion. All three possibilities can be tested with longer term multi-frequency monitoring of WX\,UMa's radio emission.
\bibliographystyle{aa}
\bibliography{ref}

\begin{thebibliography}{50}
\expandafter\ifx\csname natexlab\endcsname\relax\def\natexlab#1{#1}\fi

\bibitem[{{Alonso-Floriano} {et~al.}(2015){Alonso-Floriano}, {Morales},
  {Caballero}, {Montes}, {Klutsch}, {Mundt}, {Cort{\'e}s-Contreras}, {Ribas},
  {Reiners}, {Amado}, {Quirrenbach}, \& {Jeffers}}]{SpecClass}
{Alonso-Floriano}, F.~J., {Morales}, J.~C., {Caballero}, J.~A., {et~al.} 2015,
  \aap, 577, A128

\bibitem[{{Callingham} {et~al.}(2021{\natexlab{a}}){Callingham}, {Pope},
  {Feinstein}, {Vedantham}, {Shimwell}, {Zarka}, {Tasse}, {Lamy}, {Veken},
  {Toet}, {Sabater}, {Best}, {van Weeren}, {R{\"o}ttgering}, \&
  {Ray}}]{callingham_crdra}
{Callingham}, J.~R., {Pope}, B.~J.~S., {Feinstein}, A.~D., {et~al.}
  2021{\natexlab{a}}, arXiv e-prints, arXiv:2102.04751

\bibitem[{{Callingham} {et~al.}(2019){Callingham}, {Vedantham}, {Pope},
  {Shimwell}, \& {LoTSS Team}}]{callingham2019}
{Callingham}, J.~R., {Vedantham}, H.~K., {Pope}, B.~J.~S., {Shimwell}, T.~W.,
  \& {LoTSS Team}. 2019, Research Notes of the American Astronomical Society,
  3, 37

\bibitem[{{Callingham} {et~al.}(2021{\natexlab{b}}){Callingham}, {Vedantham},
  {Shimwell}, \& {Pope}}]{callingham2020}
{Callingham}, J.~R., {Vedantham}, H.~K., {Shimwell}, T.~W., \& {Pope}, B.~J.~S.
  2021{\natexlab{b}}, Nature Astronomy \,(under review), xxx, xxx

\bibitem[{{Cohen} {et~al.}(2014){Cohen}, {Drake}, {Glocer}, {Garraffo},
  {Poppenhaeger}, {Bell}, {Ridley}, \& {Gombosi}}]{cohen2014}
{Cohen}, O., {Drake}, J.~J., {Glocer}, A., {et~al.} 2014, \apj, 790, 57

\bibitem[{{Collier Cameron} \& {Robinson}(1989)}]{cameron1989}
{Collier Cameron}, A. \& {Robinson}, R.~D. 1989, \mnras, 238, 657

\bibitem[{{Dressing} \& {Charbonneau}(2015)}]{dressing2015}
{Dressing}, C.~D. \& {Charbonneau}, D. 2015, \apj, 807, 45

\bibitem[{{Dulk}(1985)}]{dulk1985}
{Dulk}, G.~A. 1985, \araa, 23, 169

\bibitem[{{Gaia Collaboration} {et~al.}(2018){Gaia Collaboration}, {Brown},
  {Vallenari}, {Prusti}, {de Bruijne}, {Babusiaux}, {Bailer-Jones}, {Biermann},
  {Evans}, {Eyer}, {Jansen}, {Jordi}, {Klioner}, {Lammers}, {Lindegren},
  {Luri}, {Mignard}, {Panem}, {Pourbaix}, {Randich}, {Sartoretti}, {Siddiqui},
  {Soubiran}, {van Leeuwen}, {Walton}, {Arenou}, {Bastian}, {Cropper},
  {Drimmel}, {Katz}, {Lattanzi}, {Bakker}, {Cacciari}, {Casta{\~n}eda},
  {Chaoul}, {Cheek}, {De Angeli}, {Fabricius}, {Guerra}, {Holl}, {Masana},
  {Messineo}, {Mowlavi}, {Nienartowicz}, {Panuzzo}, {Portell}, {Riello},
  {Seabroke}, {Tanga}, {Th{\'e}venin}, {Gracia-Abril}, {Comoretto},
  {Garcia-Reinaldos}, {Teyssier}, {Altmann}, {Andrae}, {Audard},
  {Bellas-Velidis}, {Benson}, {Berthier}, {Blomme}, {Burgess}, {Busso},
  {Carry}, {Cellino}, {Clementini}, {Clotet}, {Creevey}, {Davidson}, {De
  Ridder}, {Delchambre}, {Dell'Oro}, {Ducourant},
  {Fern{\'a}ndez-Hern{\'a}ndez}, {Fouesneau}, {Fr{\'e}mat}, {Galluccio},
  {Garc{\'\i}a-Torres}, {Gonz{\'a}lez-N{\'u}{\~n}ez}, {Gonz{\'a}lez-Vidal},
  {Gosset}, {Guy}, {Halbwachs}, {Hambly}, {Harrison}, {Hern{\'a}ndez},
  {Hestroffer}, {Hodgkin}, {Hutton}, {Jasniewicz}, {Jean-Antoine-Piccolo},
  {Jordan}, {Korn}, {Krone-Martins}, {Lanzafame}, {Lebzelter}, {L{\"o}ffler},
  {Manteiga}, {Marrese}, {Mart{\'\i}n-Fleitas}, {Moitinho}, {Mora}, {Muinonen},
  {Osinde}, {Pancino}, {Pauwels}, {Petit}, {Recio-Blanco}, {Richards},
  {Rimoldini}, {Robin}, {Sarro}, {Siopis}, {Smith}, {Sozzetti}, {S{\"u}veges},
  {Torra}, {van Reeven}, {Abbas}, {Abreu Aramburu}, {Accart}, {Aerts},
  {Altavilla}, {{\'A}lvarez}, {Alvarez}, {Alves}, {Anderson}, {Andrei},
  {Anglada Varela}, {Antiche}, {Antoja}, {Arcay}, {Astraatmadja}, {Bach},
  {Baker}, {Balaguer-N{\'u}{\~n}ez}, {Balm}, {Barache}, {Barata}, {Barbato},
  {Barblan}, {Barklem}, {Barrado}, {Barros}, {Barstow}, {Bartholom{\'e}
  Mu{\~n}oz}, {Bassilana}, {Becciani}, {Bellazzini}, {Berihuete}, {Bertone},
  {Bianchi}, {Bienaym{\'e}}, {Blanco-Cuaresma}, {Boch}, {Boeche}, {Bombrun},
  {Borrachero}, {Bossini}, {Bouquillon}, {Bourda}, {Bragaglia}, {Bramante},
  {Breddels}, {Bressan}, {Brouillet}, {Br{\"u}semeister}, {Brugaletta},
  {Bucciarelli}, {Burlacu}, {Busonero}, {Butkevich}, {Buzzi}, {Caffau},
  {Cancelliere}, {Cannizzaro}, {Cantat-Gaudin}, {Carballo}, {Carlucci},
  {Carrasco}, {Casamiquela}, {Castellani}, {Castro-Ginard}, {Charlot},
  {Chemin}, {Chiavassa}, {Cocozza}, {Costigan}, {Cowell}, {Crifo}, {Crosta},
  {Crowley}, {Cuypers}, {Dafonte}, {Damerdji}, {Dapergolas}, {David}, {David},
  {de Laverny}, {De Luise}, {De March}, {de Martino}, {de Souza}, {de Torres},
  {Debosscher}, {del Pozo}, {Delbo}, {Delgado}, {Delgado}, {Di Matteo},
  {Diakite}, {Diener}, {Distefano}, {Dolding}, {Drazinos}, {Dur{\'a}n},
  {Edvardsson}, {Enke}, {Eriksson}, {Esquej}, {Eynard Bontemps}, {Fabre},
  {Fabrizio}, {Faigler}, {Falc{\~a}o}, {Farr{\`a}s Casas}, {Federici},
  {Fedorets}, {Fernique}, {Figueras}, {Filippi}, {Findeisen}, {Fonti},
  {Fraile}, {Fraser}, {Fr{\'e}zouls}, {Gai}, {Galleti}, {Garabato},
  {Garc{\'\i}a-Sedano}, {Garofalo}, {Garralda}, {Gavel}, {Gavras}, {Gerssen},
  {Geyer}, {Giacobbe}, {Gilmore}, {Girona}, {Giuffrida}, {Glass}, {Gomes},
  {Granvik}, {Gueguen}, {Guerrier}, {Guiraud}, {Guti{\'e}rrez-S{\'a}nchez},
  {Haigron}, {Hatzidimitriou}, {Hauser}, {Haywood}, {Heiter}, {Helmi}, {Heu},
  {Hilger}, {Hobbs}, {Hofmann}, {Holland}, {Huckle}, {Hypki}, {Icardi},
  {Jan{\ss}en}, {Jevardat de Fombelle}, {Jonker}, {Juh{\'a}sz}, {Julbe},
  {Karampelas}, {Kewley}, {Klar}, {Kochoska}, {Kohley}, {Kolenberg},
  {Kontizas}, {Kontizas}, {Koposov}, {Kordopatis}, {Kostrzewa-Rutkowska},
  {Koubsky}, {Lambert}, {Lanza}, {Lasne}, {Lavigne}, {Le Fustec}, {Le
  Poncin-Lafitte}, {Lebreton}, {Leccia}, {Leclerc}, {Lecoeur-Taibi},
  {Lenhardt}, {Leroux}, {Liao}, {Licata}, {Lindstr{\o}m}, {Lister}, {Livanou},
  {Lobel}, {L{\'o}pez}, {Managau}, {Mann}, {Mantelet}, {Marchal}, {Marchant},
  {Marconi}, {Marinoni}, {Marschalk{\'o}}, {Marshall}, {Martino}, {Marton},
  {Mary}, {Massari}, {Matijevi{\v{c}}}, {Mazeh}, {McMillan}, {Messina},
  {Michalik}, {Millar}, {Molina}, {Molinaro}, {Moln{\'a}r}, {Montegriffo},
  {Mor}, {Morbidelli}, {Morel}, {Morris}, {Mulone}, {Muraveva}, {Musella},
  {Nelemans}, {Nicastro}, {Noval}, {O'Mullane}, {Ord{\'e}novic},
  {Ord{\'o}{\~n}ez-Blanco}, {Osborne}, {Pagani}, {Pagano}, {Pailler},
  {Palacin}, {Palaversa}, {Panahi}, {Pawlak}, {Piersimoni}, {Pineau}, {Plachy},
  {Plum}, {Poggio}, {Poujoulet}, {Pr{\v{s}}a}, {Pulone}, {Racero}, {Ragaini},
  {Rambaux}, {Ramos-Lerate}, {Regibo}, {Reyl{\'e}}, {Riclet}, {Ripepi}, {Riva},
  {Rivard}, {Rixon}, {Roegiers}, {Roelens}, {Romero-G{\'o}mez}, {Rowell},
  {Royer}, {Ruiz-Dern}, {Sadowski}, {Sagrist{\`a} Sell{\'e}s}, {Sahlmann},
  {Salgado}, {Salguero}, {Sanna}, {Santana-Ros}, {Sarasso}, {Savietto},
  {Schultheis}, {Sciacca}, {Segol}, {Segovia}, {S{\'e}gransan}, {Shih},
  {Siltala}, {Silva}, {Smart}, {Smith}, {Solano}, {Solitro}, {Sordo}, {Soria
  Nieto}, {Souchay}, {Spagna}, {Spoto}, {Stampa}, {Steele},
  {Steidelm{\"u}ller}, {Stephenson}, {Stoev}, {Suess}, {Surdej}, {Szabados},
  {Szegedi-Elek}, {Tapiador}, {Taris}, {Tauran}, {Taylor}, {Teixeira},
  {Terrett}, {Teyssand ier}, {Thuillot}, {Titarenko}, {Torra Clotet}, {Turon},
  {Ulla}, {Utrilla}, {Uzzi}, {Vaillant}, {Valentini}, {Valette}, {van Elteren},
  {Van Hemelryck}, {van Leeuwen}, {Vaschetto}, {Vecchiato}, {Veljanoski},
  {Viala}, {Vicente}, {Vogt}, {von Essen}, {Voss}, {Votruba}, {Voutsinas},
  {Walmsley}, {Weiler}, {Wertz}, {Wevers}, {Wyrzykowski}, {Yoldas},
  {{\v{Z}}erjal}, {Ziaeepour}, {Zorec}, {Zschocke}, {Zucker}, {Zurbach}, \&
  {Zwitter}}]{gaia}
{Gaia Collaboration}, {Brown}, A.~G.~A., {Vallenari}, A., {et~al.} 2018, \aap,
  616, A1

\bibitem[{{Garraffo} {et~al.}(2017){Garraffo}, {Drake}, {Cohen},
  {Alvarado-G{\'o}mez}, \& {Moschou}}]{garraffo2017}
{Garraffo}, C., {Drake}, J.~J., {Cohen}, O., {Alvarado-G{\'o}mez}, J.~D., \&
  {Moschou}, S.~P. 2017, \apjl, 843, L33

\bibitem[{{Gillon} {et~al.}(2017){Gillon}, {Triaud}, {Demory}, {Jehin}, {Agol},
  {Deck}, {Lederer}, {de Wit}, {Burdanov}, {Ingalls}, {Bolmont}, {Leconte},
  {Raymond}, {Selsis}, {Turbet}, {Barkaoui}, {Burgasser}, {Burleigh}, {Carey},
  {Chaushev}, {Copperwheat}, {Delrez}, {Fernand es}, {Holdsworth}, {Kotze},
  {Van Grootel}, {Almleaky}, {Benkhaldoun}, {Magain}, \& {Queloz}}]{gillon2017}
{Gillon}, M., {Triaud}, A. H.~M.~J., {Demory}, B.-O., {et~al.} 2017, \nat, 542,
  456

\bibitem[{{Goldreich} \& {Lynden-Bell}(1969)}]{goldreich1969}
{Goldreich}, P. \& {Lynden-Bell}, D. 1969, \apj, 156, 59

\bibitem[{{G{\"u}del}(2002)}]{gudel2002}
{G{\"u}del}, M. 2002, \araa, 40, 217

\bibitem[{{Hancock} {et~al.}(2012){Hancock}, {Murphy}, {Gaensler}, {Hopkins},
  \& {Curran}}]{2012MNRAS.422.1812H}
{Hancock}, P.~J., {Murphy}, T., {Gaensler}, B.~M., {Hopkins}, A., \& {Curran},
  J.~R. 2012, \mnras, 422, 1812

\bibitem[{{Havnes} \& {Goertz}(1984)}]{havnes1984}
{Havnes}, O. \& {Goertz}, C.~K. 1984, \aap, 138, 421

\bibitem[{{Hewitt} {et~al.}(1982){Hewitt}, {Melrose}, \&
  {Ronnmark}}]{hewitt1982}
{Hewitt}, R.~G., {Melrose}, D.~B., \& {Ronnmark}, K.~G. 1982, Australian
  Journal of Physics, 35, 447

\bibitem[{{Jardine} {et~al.}(2020){Jardine}, {Collier Cameron}, {Donati}, \&
  {Hussain}}]{jardine2020}
{Jardine}, M., {Collier Cameron}, A., {Donati}, J.~F., \& {Hussain}, G.~A.~J.
  2020, \mnras, 491, 4076

\bibitem[{{Johnstone} \& {G{\"u}del}(2015)}]{johnstone2015}
{Johnstone}, C.~P. \& {G{\"u}del}, M. 2015, \aap, 578, A129

\bibitem[{{Kavanagh} {et~al.}(2021){Kavanagh}, {Vidotto}, {Klein}, {Jardine},
  {Donati}, \& {Fionnag{\'a}in}}]{Kavanagh2021}
{Kavanagh}, R.~D., {Vidotto}, A.~A., {Klein}, B., {et~al.} 2021, \mnras
  [\eprint[arXiv]{2103.16318}]

\bibitem[{{Kopparapu} {et~al.}(2013){Kopparapu}, {Ramirez}, {Kasting}, {Eymet},
  {Robinson}, {Mahadevan}, {Terrien}, {Domagal-Goldman}, {Meadows}, \&
  {Deshpande}}]{kopparapu2013}
{Kopparapu}, R.~K., {Ramirez}, R., {Kasting}, J.~F., {et~al.} 2013, \apj, 765,
  131

\bibitem[{{Lacy} {et~al.}(2020){Lacy}, {Baum}, {Chandler}, {Chatterjee},
  {Clarke}, {Deustua}, {English}, {Farnes}, {Gaensler}, {Gugliucci},
  {Hallinan}, {Kent}, {Kimball}, {Law}, {Lazio}, {Marvil}, {Mao}, {Medlin},
  {Mooley}, {Murphy}, {Myers}, {Osten}, {Richards}, {Rosolowsky}, {Rudnick},
  {Schinzel}, {Sivakoff}, {Sjouwerman}, {Taylor}, {White}, {Wrobel},
  {Andernach}, {Beasley}, {Berger}, {Bhatnager}, {Birkinshaw}, {Bower},
  {Brandt}, {Brown}, {Burke-Spolaor}, {Butler}, {Comerford}, {Demorest}, {Fu},
  {Giacintucci}, {Golap}, {G{\"u}th}, {Hales}, {Hiriart}, {Hodge}, {Horesh},
  {Ivezi{\'c}}, {Jarvis}, {Kamble}, {Kassim}, {Liu}, {Loinard}, {Lyons},
  {Masters}, {Mezcua}, {Moellenbrock}, {Mroczkowski}, {Nyland}, {O'Dea},
  {O'Sullivan}, {Peters}, {Radford}, {Rao}, {Robnett}, {Salcido}, {Shen},
  {Sobotka}, {Witz}, {Vaccari}, {van Weeren}, {Vargas}, {Williams}, \&
  {Yoon}}]{vlass}
{Lacy}, M., {Baum}, S.~A., {Chandler}, C.~J., {et~al.} 2020, \pasp, 132, 035001

\bibitem[{{Linsky} {et~al.}(1992){Linsky}, {Drake}, \& {Bastian}}]{linsky1992}
{Linsky}, J.~L., {Drake}, S.~A., \& {Bastian}, T.~S. 1992, \apj, 393, 341

\bibitem[{{Litvinenko} {et~al.}(2009){Litvinenko}, {Lecacheux}, {Rucker},
  {Konovalenko}, {Ryabov}, {Taubenschuss}, {Vinogradov}, \&
  {Shaposhnikov}}]{litvinenko2009}
{Litvinenko}, G.~V., {Lecacheux}, A., {Rucker}, H.~O., {et~al.} 2009, \aap,
  493, 651

\bibitem[{{Lynch} {et~al.}(2017){Lynch}, {Lenc}, {Kaplan}, {Murphy}, \&
  {Anderson}}]{lynch2017}
{Lynch}, C.~R., {Lenc}, E., {Kaplan}, D.~L., {Murphy}, T., \& {Anderson}, G.~E.
  2017, \apjl, 836, L30

\bibitem[{{Melrose} \& {Dulk}(1982)}]{melrose1982}
{Melrose}, D.~B. \& {Dulk}, G.~A. 1982, \apj, 259, 844

\bibitem[{{Morin} {et~al.}(2010){Morin}, {Donati}, {Petit}, {Delfosse},
  {Forveille}, \& {Jardine}}]{morin2010}
{Morin}, J., {Donati}, J.~F., {Petit}, P., {et~al.} 2010, \mnras, 407, 2269

\bibitem[{{Newton} {et~al.}(2017){Newton}, {Irwin}, {Charbonneau}, {Berlind},
  {Calkins}, \& {Mink}}]{MassRef}
{Newton}, E.~R., {Irwin}, J., {Charbonneau}, D., {et~al.} 2017, \apj, 834, 85

\bibitem[{{Pneuman}(1968)}]{pneuman1968}
{Pneuman}, G.~W. 1968, \solphys, 3, 578

\bibitem[{{Priest} \& {Forbes}(2002)}]{priest2002}
{Priest}, E.~R. \& {Forbes}, T.~G. 2002, \aapr, 10, 313

\bibitem[{{Shimwell} {et~al.}(2017){Shimwell}, {R{\"o}ttgering}, {Best},
  {Williams}, {Dijkema}, {de Gasperin}, {Hardcastle}, {Heald}, {Hoang},
  {Horneffer}, {Intema}, {Mahony}, {Mandal}, {Mechev}, {Morabito}, {Oonk},
  {Rafferty}, {Retana-Montenegro}, {Sabater}, {Tasse}, {van Weeren},
  {Br{\"u}ggen}, {Brunetti}, {Chy{\.z}y}, {Conway}, {Haverkorn}, {Jackson},
  {Jarvis}, {McKean}, {Miley}, {Morganti}, {White}, {Wise}, {van Bemmel},
  {Beck}, {Brienza}, {Bonafede}, {Calistro Rivera}, {Cassano}, {Clarke},
  {Cseh}, {Deller}, {Drabent}, {van Driel}, {Engels}, {Falcke}, {Ferrari},
  {Fr{\"o}hlich}, {Garrett}, {Harwood}, {Heesen}, {Hoeft}, {Horellou},
  {Israel}, {Kapi{\'n}ska}, {Kunert-Bajraszewska}, {McKay}, {Mohan},
  {Orr{\'u}}, {Pizzo}, {Prandoni}, {Schwarz}, {Shulevski}, {Sipior}, {Smith},
  {Sridhar}, {Steinmetz}, {Stroe}, {Varenius}, {van der Werf}, {Zensus}, \&
  {Zwart}}]{lotss1}
{Shimwell}, T.~W., {R{\"o}ttgering}, H.~J.~A., {Best}, P.~N., {et~al.} 2017,
  \aap, 598, A104

\bibitem[{{Shimwell} {et~al.}(2019){Shimwell}, {Tasse}, {Hardcastle}, {Mechev},
  {Williams}, {Best}, {R{\"o}ttgering}, {Callingham}, {Dijkema}, {de Gasperin},
  {Hoang}, {Hugo}, {Mirmont}, {Oonk}, {Prandoni}, {Rafferty}, {Sabater},
  {Smirnov}, {van Weeren}, {White}, {Atemkeng}, {Bester}, {Bonnassieux},
  {Br{\"u}ggen}, {Brunetti}, {Chy{\.z}y}, {Cochrane}, {Conway}, {Croston},
  {Danezi}, {Duncan}, {Haverkorn}, {Heald}, {Iacobelli}, {Intema}, {Jackson},
  {Jamrozy}, {Jarvis}, {Lakhoo}, {Mevius}, {Miley}, {Morabito}, {Morganti},
  {Nisbet}, {Orr{\'u}}, {Perkins}, {Pizzo}, {Schrijvers}, {Smith}, {Vermeulen},
  {Wise}, {Alegre}, {Bacon}, {van Bemmel}, {Beswick}, {Bonafede}, {Botteon},
  {Bourke}, {Brienza}, {Calistro Rivera}, {Cassano}, {Clarke}, {Conselice},
  {Dettmar}, {Drabent}, {Dumba}, {Emig}, {En{\ss}lin}, {Ferrari}, {Garrett},
  {G{\'e}nova-Santos}, {Goyal}, {G{\"u}rkan}, {Hale}, {Harwood}, {Heesen},
  {Hoeft}, {Horellou}, {Jackson}, {Kokotanekov}, {Kondapally},
  {Kunert-Bajraszewska}, {Mahatma}, {Mahony}, {Mandal}, {McKean}, {Merloni},
  {Mingo}, {Miskolczi}, {Mooney}, {Nikiel-Wroczy{\'n}ski}, {O'Sullivan},
  {Quinn}, {Reich}, {Roskowi{\'n}ski}, {Rowlinson}, {Savini}, {Saxena},
  {Schwarz}, {Shulevski}, {Sridhar}, {Stacey}, {Urquhart}, {van der Wiel},
  {Varenius}, {Webster}, \& {Wilber}}]{lotss2}
{Shimwell}, T.~W., {Tasse}, C., {Hardcastle}, M.~J., {et~al.} 2019, \aap, 622,
  A1

\bibitem[{{Stepanov} {et~al.}(2001){Stepanov}, {Kliem}, {Zaitsev}, {F{\"u}rst},
  {Jessner}, {Kr{\"u}ger}, {Hildebrand t}, \& {Schmitt}}]{stepanov2001}
{Stepanov}, A.~V., {Kliem}, B., {Zaitsev}, V.~V., {et~al.} 2001, \aap, 374,
  1072

\bibitem[{{Tasse} {et~al.}(2020){Tasse}, {Shimwell}, {Hardcastle},
  {O'Sullivan}, {van Weeren}, {Best}, {Bester}, {Hugo}, {Smirnov}, {Sabater},
  {Calistro-Rivera}, {de Gasperin}, {Morabito}, {R{\"o}ttgering}, {Williams},
  {Bonato}, {Bondi}, {Botteon}, {Br{\"u}ggen}, {Brunetti}, {Chy{\.z}y},
  {Garrett}, {G{\"u}rkan}, {Jarvis}, {Kondapally}, {Mandal}, {Prandoni},
  {Repetti}, {Retana-Montenegro}, {Schwarz}, {Shulevski}, \&
  {Wiaux}}]{tasse2020}
{Tasse}, C., {Shimwell}, T., {Hardcastle}, M.~J., {et~al.} 2020, arXiv
  e-prints, arXiv:2011.08328

\bibitem[{{Treumann}(2006)}]{treumann}
{Treumann}, R.~A. 2006, \aapr, 13, 229

\bibitem[{{Trigilio} {et~al.}(2000){Trigilio}, {Leto}, {Leone}, {Umana}, \&
  {Buemi}}]{triglio2000}
{Trigilio}, C., {Leto}, P., {Leone}, F., {Umana}, G., \& {Buemi}, C. 2000,
  \aap, 362, 281

\bibitem[{{van Haarlem} {et~al.}(2013){van Haarlem}, {Wise}, {Gunst}, {Heald},
  {McKean}, {Hessels}, {de Bruyn}, {Nijboer}, {Swinbank}, {Fallows},
  {Brentjens}, {Nelles}, {Beck}, {Falcke}, {Fender}, {H{\"o}randel},
  {Koopmans}, {Mann}, {Miley}, {R{\"o}ttgering}, {Stappers}, {Wijers},
  {Zaroubi}, {van den Akker}, {Alexov}, {Anderson}, {Anderson}, {van Ardenne},
  {Arts}, {Asgekar}, {Avruch}, {Batejat}, {B{\"a}hren}, {Bell}, {Bell}, {van
  Bemmel}, {Bennema}, {Bentum}, {Bernardi}, {Best}, {B{\^\i}rzan}, {Bonafede},
  {Boonstra}, {Braun}, {Bregman}, {Breitling}, {van de Brink}, {Broderick},
  {Broekema}, {Brouw}, {Br{\"u}ggen}, {Butcher}, {van Cappellen}, {Ciardi},
  {Coenen}, {Conway}, {Coolen}, {Corstanje}, {Damstra}, {Davies}, {Deller},
  {Dettmar}, {van Diepen}, {Dijkstra}, {Donker}, {Doorduin}, {Dromer}, {Drost},
  {van Duin}, {Eisl{\"o}ffel}, {van Enst}, {Ferrari}, {Frieswijk}, {Gankema},
  {Garrett}, {de Gasperin}, {Gerbers}, {de Geus}, {Grie{\ss}meier}, {Grit},
  {Gruppen}, {Hamaker}, {Hassall}, {Hoeft}, {Holties}, {Horneffer}, {van der
  Horst}, {van Houwelingen}, {Huijgen}, {Iacobelli}, {Intema}, {Jackson},
  {Jelic}, {de Jong}, {Juette}, {Kant}, {Karastergiou}, {Koers}, {Kollen},
  {Kondratiev}, {Kooistra}, {Koopman}, {Koster}, {Kuniyoshi}, {Kramer},
  {Kuper}, {Lambropoulos}, {Law}, {van Leeuwen}, {Lemaitre}, {Loose}, {Maat},
  {Macario}, {Markoff}, {Masters}, {McFadden}, {McKay-Bukowski}, {Meijering},
  {Meulman}, {Mevius}, {Middelberg}, {Millenaar}, {Miller-Jones}, {Mohan},
  {Mol}, {Morawietz}, {Morganti}, {Mulcahy}, {Mulder}, {Munk}, {Nieuwenhuis},
  {van Nieuwpoort}, {Noordam}, {Norden}, {Noutsos}, {Offringa}, {Olofsson},
  {Omar}, {Orr{\'u}}, {Overeem}, {Paas}, {Pand ey-Pommier}, {Pandey}, {Pizzo},
  {Polatidis}, {Rafferty}, {Rawlings}, {Reich}, {de Reijer}, {Reitsma},
  {Renting}, {Riemers}, {Rol}, {Romein}, {Roosjen}, {Ruiter}, {Scaife}, {van
  der Schaaf}, {Scheers}, {Schellart}, {Schoenmakers}, {Schoonderbeek},
  {Serylak}, {Shulevski}, {Sluman}, {Smirnov}, {Sobey}, {Spreeuw}, {Steinmetz},
  {Sterks}, {Stiepel}, {Stuurwold}, {Tagger}, {Tang}, {Tasse}, {Thomas},
  {Thoudam}, {Toribio}, {van der Tol}, {Usov}, {van Veelen}, {van der Veen},
  {ter Veen}, {Verbiest}, {Vermeulen}, {Vermaas}, {Vocks}, {Vogt}, {de Vos},
  {van der Wal}, {van Weeren}, {Weggemans}, {Weltevrede}, {White}, {Wijnholds},
  {Wilhelmsson}, {Wucknitz}, {Yatawatta}, {Zarka}, {Zensus}, \& {van
  Zwieten}}]{lofar}
{van Haarlem}, M.~P., {Wise}, M.~W., {Gunst}, A.~W., {et~al.} 2013, \aap, 556,
  A2

\bibitem[{{van Straten} {et~al.}(2010){van Straten}, {Manchester}, {Johnston},
  \& {Reynolds}}]{vanstraten}
{van Straten}, W., {Manchester}, R.~N., {Johnston}, S., \& {Reynolds}, J.~E.
  2010, \pasa, 27, 104

\bibitem[{{van Weeren} {et~al.}(2020){van Weeren}, {Shimwell}, {Botteon},
  {Brunetti}, {Br{\"u}ggen}, {Boxelaar}, {Cassano}, {Di Gennaro},
  {Andrade-Santos}, {Bonnassieux}, {Bonafede}, {Cuciti}, {Dallacasa}, {de
  Gasperin}, {Gastaldello}, {Hardcastle}, {Hoeft}, {Kraft}, {Mandal},
  {Rossetti}, {R{\"o}ttgering}, {Tasse}, \& {Wilber}}]{2020arXiv201102387V}
{van Weeren}, R.~J., {Shimwell}, T.~W., {Botteon}, A., {et~al.} 2020, arXiv
  e-prints, arXiv:2011.02387

\bibitem[{{Vedantham}(2020)}]{vedantham2020b}
{Vedantham}, H.~K. 2020, arXiv e-prints, arXiv:2008.05707

\bibitem[{{Vedantham} {et~al.}(2020{\natexlab{a}}){Vedantham}, {Callingham},
  {Shimwell}, {Dupuy}, {Best}, {Liu}, {Zhang}, {De}, {Lamy}, {Zarka},
  {R{\"o}ttgering}, \& {Shulevski}}]{elegast}
{Vedantham}, H.~K., {Callingham}, J.~R., {Shimwell}, T.~W., {et~al.}
  2020{\natexlab{a}}, \apjl, 903, L33

\bibitem[{{Vedantham} {et~al.}(2020{\natexlab{b}}){Vedantham}, {Callingham},
  {Shimwell}, {Dupuy}, {Best}, {Liu}, {Zhang}, {De}, {Lamy}, {Zarka},
  {Rottgering}, \& {Shulevski}}]{vedantham-bdc}
{Vedantham}, H.~K., {Callingham}, J.~R., {Shimwell}, T.~W., {et~al.}
  2020{\natexlab{b}}, arXiv e-prints, arXiv:2010.01915

\bibitem[{{Vedantham} {et~al.}(2020{\natexlab{c}}){Vedantham}, {Callingham},
  {Shimwell}, {Tasse}, {Pope}, {Bedell}, {Snellen}, {Best}, {Hardcastle},
  {Haverkorn}, {Mechev}, {O'Sullivan}, {R{\"o}ttgering}, \&
  {White}}]{vedantham2020}
{Vedantham}, H.~K., {Callingham}, J.~R., {Shimwell}, T.~W., {et~al.}
  2020{\natexlab{c}}, Nature Astronomy, 4, 577

\bibitem[{{Vidotto}(2021)}]{vidotto2021}
{Vidotto}, A.~A. 2021, In press, Living Reviews in Solar Physics

\bibitem[{{Vidotto} {et~al.}(2013){Vidotto}, {Jardine}, {Morin}, {Donati},
  {Lang}, \& {Russell}}]{vidotto2013}
{Vidotto}, A.~A., {Jardine}, M., {Morin}, J., {et~al.} 2013, \aap, 557, A67

\bibitem[{{Vidotto} {et~al.}(2014){Vidotto}, {Jardine}, {Morin}, {Donati},
  {Opher}, \& {Gombosi}}]{vidotto2014}
{Vidotto}, A.~A., {Jardine}, M., {Morin}, J., {et~al.} 2014, \mnras, 438, 1162

\bibitem[{{Villadsen} \& {Hallinan}(2019)}]{villadsen2019}
{Villadsen}, J. \& {Hallinan}, G. 2019, \apj, 871, 214

\bibitem[{{Wild} \& {Smerd}(1972)}]{wild1972}
{Wild}, J.~P. \& {Smerd}, S.~F. 1972, \araa, 10, 159

\bibitem[{{Zaitsev} \& {Stepanov}(1983)}]{zaitsev1983}
{Zaitsev}, V.~V. \& {Stepanov}, A.~V. 1983, \solphys, 88, 297

\bibitem[{{Zarka}(1998)}]{zarka1998}
{Zarka}, P. 1998, \jgr, 103, 20159

\bibitem[{{Zheleznyakov}(1996)}]{1996ASSL..204.....Z}
{Zheleznyakov}, V.~V. 1996, {Radiation in Astrophysical Plasmas}, Vol. 204
  ({Kluwer academic publishers})

\end{thebibliography}

\begin{acknowledgements}
ID completed this work as part of the ASTRON summer student programme. JRC thanks the Nederlandse Organisatie voor Wetenschappelijk Onderzoek (NWO) for support via the Talent Programme Veni grant.
AAV and TPR acknowledge funding from the European Research Council (ERC) under
the European Union’s Horizon 2020 Research and Innovation Programme
(grant agreements No 817540, ASTROFLOW and No 743029 EASY respectively).
AD acknowledges support by the BMBF Verbundforschung under the grant 05A20STA.
This paper is based on data obtained with the International LOFAR Telescope (obs. IDs 233996, 441476, 403966 and 403968) as part of the LoTSS survey. LOFAR is the Low Frequency Array designed and constructed by ASTRON. It has observing, data processing, and data storage facilities in several countries, that are owned by various parties (each with their own funding sources), and that are collectively operated by the ILT foundation under a joint scientific policy. The ILT resources have benefitted from the following recent major funding sources: CNRS-INSU, Observatoire de Paris and Université d'Orléans, France; BMBF, MIWF-NRW, MPG, Germany; Science Foundation Ireland (SFI), Department of Business, Enterprise and Innovation (DBEI), Ireland; NWO, The Netherlands; The Science and Technology Facilities Council, UK.
This research made use of the Dutch national e-infrastructure with support of the SURF Cooperative (e-infra 180169) and the LOFAR e-infra group. The J\"{u}lich LOFAR Long Term Archive and the German LOFAR network are both coordinated and operated by the J\"{u}lich Supercomputing Centre (JSC), and computing resources on the supercomputer JUWELS at JSC were provided by the Gauss Centre for supercomputing e.V. (grant CHTB00) through the John von Neumann Institute for Computing (NIC).
This research made use of the University of Hertfordshire high-performance computing facility and the LOFAR-UK computing facility located at the University of Hertfordshire and supported by STFC [ST/P000096/1], and of the Italian LOFAR IT computing infrastructure supported and operated by INAF, and by the Physics Department of Turin university (under an agreement with Consorzio Interuniversitario per la Fisica Spaziale) at the C3S Supercomputing Centre, Italy.
The J\"{u}lich LOFAR Long Term Archive and the German LOFAR network are both coordinated and operated by the J\"{u}lich Supercomputing Centre (JSC), and computing resources on the supercomputer JUWELS at JSC were provided by the Gauss Centre for Supercomputing e.V. (grant CHTB00) through the John von Neumann Institute for Computing (NIC).
\end{acknowledgements}

\begin{appendix}
\section{Radio data processing}
\label{sec:data_proc}
The data were processed with the usual LoTSS pipeline\footnote{The LoTSS pipeline can be found at https://github.com/mhardcastle/ddf-pipeline} \citep{lotss1,lotss2} which is described in \cite{tasse2020}. We then performed an additional round of self-calibration in the direction of WX\,UMa with a procedure described in \cite{2020arXiv201102387V} and also employed in \citep{vedantham2020}.
After calibration, the measurement set was further processed to make source extraction easier for producing light curves and spectra.  WX\,UMa was removed from the mask image which was then used to produce a model file using \texttt{WSClean}.  WX\,UMa was also removed from these model images so that it would not be removed during the subtraction of other sources.  The subtracted model image was then used with \texttt{WSClean} again to predict visibilities which were recorded in a new modelled data column in the measurement set.  The modelled data was subtracted from the real data column and recorded in a new column - this served to remove all other bright sources from the region besides WX\,UMa. Finally, \texttt{DPPP} was used to phaseshift the measurement sets to be centred on WX\,UMa, using the subtracted data as the new real data.

After extra sources were removed from the field and the set was phaseshifted, frequency and time slices for Stokes-I and V were produced via \texttt{WSClean}. Because the signal to noise ratio of the observation was not large enough to produce a full dynamic spectrum, the frequency slices were averaged over all time and the time slices were averaged over the first 40 channels (120 to to 135.83\,MHz, $\approx$16 MHz bandwidth) where WX\,UMa is the brightest. The Background and Noise Estimation (\texttt{BANE}) tool in \texttt{Aegean} \citep{2012MNRAS.422.1812H} was used with each image to produce rms and background images. \texttt{Aegean} was run with a 4$\sigma$ seedclip to avoid identifying extraneous sources. The flux density for WX\,UMa for a particular image was assigned from the peak flux found by \texttt{Aegean} if a source was found in the expected location of WX\,UMa, and the rms noise was retrieved from the rms image produced by \texttt{BANE}. For images where no 4$\,\sigma$ source was found at the star's coordinates, a priorized fitting was applied using the catalogue from the image with the most significant detection in order to extract $\approx 3\,\sigma$ detections at the expected position of WX\,UMa. The images were also inspected by eye to make sure the results of \texttt{Aegean} were consistent with the images. The temporal evolution of WX\,UMa's radio flux is shown in Fig. \ref{fig:lc} and its spectral evolution is shown in Fig. \ref{fig:spec3}.

We note that the some the images in the light curve estimation have non-Gaussian noise due to residual sidelobe noise (from poorer $uv$ coverage) and low-level calibration errors.

\subsection{Stokes-V sign convention}\label{app:sign_convention}
There are two commonly used Stokes-V sign conventions in literature. The pulsar literature typically follows the convention ${\rm V} = {\rm LCP}- {\rm RCP}$ whereas the IAU convention is $V = {\rm RCP} - {\rm LCP}$ \citep{vanstraten}. To check the convention applicable to the LoTSS images, we compared the sign of Stokes-V of several pulsar detections to that in the literature and found that the LoTSS survey data products follow the same convention as the one in pulsar literature \citep{callingham2020}. Hence, WX\,UMa is radiating ${\rm RCP}$ radiation, which means that the electric field vector rotates counter-clockwise as a function of time to an observer looking at incoming radiation.

\section{Emission mechanism}\label{app:emission_mechanism}
\subsection{Brightness temperature of plasma emission}
\begin{figure}
    \centering
    \includegraphics[width=\linewidth]{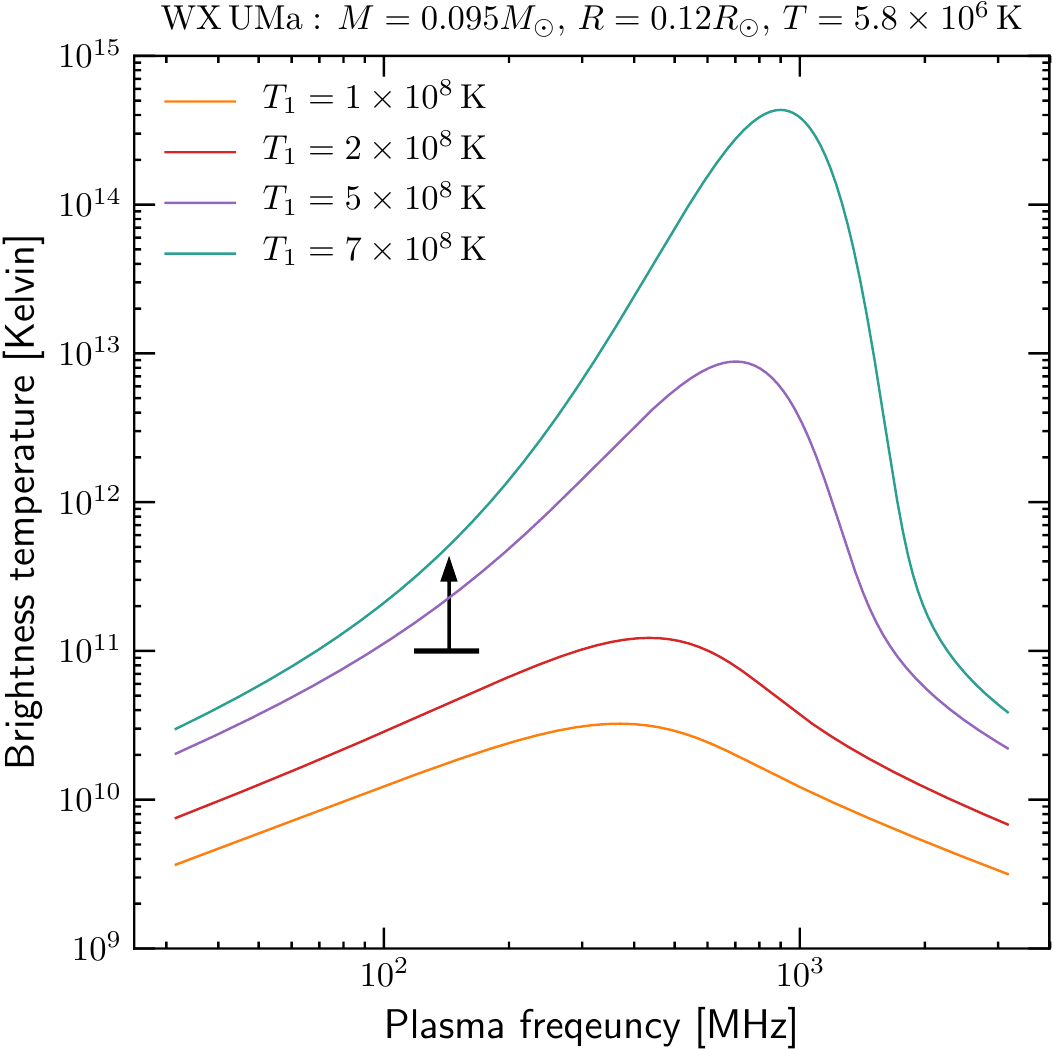}
    \caption{Brightness temperature of plasma emission computed for parameters relevant for WX\,UMa (annotated in the figure). The black bar is the measured value based on the radiation escape conditions and the assumption that the entire projected coronal surface is radiating.}
    \label{fig:plasma_tb}
\end{figure}

In Fig. \ref{fig:plasma_tb}, we computed the brightness temperature of plasma emission using the expressions of \citet[][speficially, their eqn. 15]{stepanov2001}. In doing so, we assumed a peak turbulence level of $w=10^{-5}$ and to compute the plasma temperature, we used the empirical relationship of \citet{johnstone2015} and the observed X-ray luminosity (Table \ref{tab:lit_values}).
\subsection{Escape of plasma emission}
Fundamental plasma emission is excited at the upper-hybrid frequency: $\nu_{\rm uh} \approx \sqrt{\nu_p^2+\nu_c^2}$ where $\nu_p$ and $\nu_c$ are the plasma and cyclotron frequencies respectively. Note that the upper-hybrid frequency is always larger than the cyclotron frequency. 
The cyclotron absorption optical depth for  $o$-mode radiation propagating at an angle $\theta$ to the magnetic field may be approximated by 
\begin{equation}
    \tau = \left(\frac{\pi}{2}\right)^{5/2}\frac{2}{c}\frac{\nu_p^2}{\nu}\frac{s^{2s}}{s!}\left(\frac{k_BT\sin^2\theta}{2m_ec^2} \right)^{s-1} (1+|\cos\theta|)^2 L_B
    \label{eqn:cycl_abs}
\end{equation}
where integer $s=\nu/\nu_c>1$ is the cyclotron harmonic number, $T$ is the coronal temperature, $m_e$ is the electron mass, $c$ is the speed of light, $k_B$ is Boltzmann's constant and $L_B$ is the magnetic scale length which we take to be $R_\ast$. For the parameters of interest: $T\sim 10^{6.7}\,{\rm K}$, $L_B\sim 10^{10}\,{\rm cm}$, the optical depth is safely below unity only for $s\ge4$. Hence, the radiation escape condition is $\sqrt{\nu_p^2+\nu_c^2}> 4\nu_c$, which gives $\nu_c<0.26\nu_p$.

\section{Spectral evolution of the emission-cone}\label{app:SpectralEvolution}
\begin{figure}
    \centering
    \includegraphics[width=\linewidth]{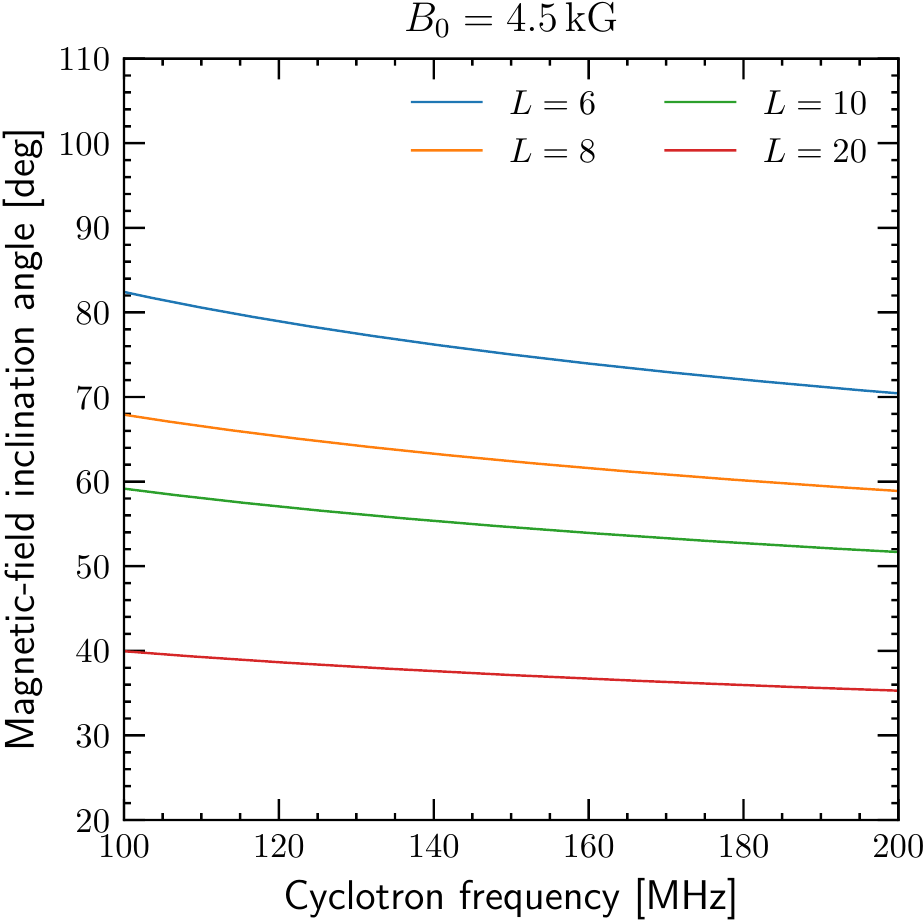}
    \caption{Polar angle of the magnetic field at the site of emission (y-axis) versus frequency of emission (x-axis) for different field lines specified by the radial distance at which they intersect the equatorial plane.}
    \label{fig:alpha_vs_nu}
\end{figure}

Consider a single dipolar field line which intersects the equatorial plane at radial distance $L$. The radial distance and polar angle, $\theta^\prime$ along the field line are related via $L = R\sin^2\theta^\prime$. The polar inclination angle of the field lines, $
\alpha$ is given by $\cos\alpha = (3\cos^2\theta^\prime-1)(1+3\cos^2\theta^\prime)^{-1/2}$.  The magnetic field strength (and cyclotron frequency) vary continuously along the field line according to $\nu(R) = \nu_0R^3_\ast/(2R^3)(4-3R/L)^{1/2}$, where $\nu_0$ is the peak cyclotron frequency at $\theta^\prime=0,\,R=R_\ast$. With the above relationships, we predicted the inclination of the emission cone axis for various $L$-shell at a range of cyclotron frequencies (see Fig. \ref{fig:alpha_vs_nu}).  The numerically computed slope of the $\alpha$ versus $\nu$ lines at $140\,{\rm MHz}$ vary between $\partial\alpha/\partial\nu\approx 0.2^\circ/{\rm MHz}$ and $\partial\alpha/\partial\nu\approx 0.8^\circ/{\rm MHz}$ as $L$ is varied between $6$ and $20$. The magnitude of the slope decreases for larger $L$ and must asymptote to zero as $L$ tends to infinity. 
\end{appendix}
\end{document}